\documentclass{revtex4}
\usepackage{graphicx,amsmath,amssymb,mathrsfs}
\usepackage{epsfig}

\begin{document}
\title{Currents in complex polymers: an example of superstatistics
for short time series}
\author{ G.Cigdem Yalcin$^{1}$\footnote{
corresponding author, phone +90 212 455 57 00 ext 15489,
Fax +90 212 455 5855, gcyalcin@istanbul.edu.tr}
 and Christian Beck$^{2}$}
\affiliation{$^{1}$Department of Physics, Istanbul University,  34134,  Vezneciler, Istanbul, Turkey
\\
$^{2}$School of Mathematical Sciences, Queen Mary, University of London, London E1 4NS, UK}

\begin{abstract}
We apply superstatistical techniques to an experimental time series
of measured transient currents through a thin Aluminium-PMMA-Aluminium
film. We show that in good approximation the current can
be approximated by local Gaussian processes with fluctuating
variance. The marginal density exhibits 'fat tails' and is
well modelled by a superstatistical
model. Our techniques can be generally be applied to other
short time series as well.
\end{abstract}

\vspace{2cm}

\maketitle

\section{Introduction}

Many time series generated by complex systems in nature are
superstatistical, i.e  they consist
of a superposition of several dynamics on
well-separated
time scales. Often, there is locally
a simple dynamics (for example, a Gaussian
process) and the parameters of that simple process fluctuate on
a much larger scale.
Such varying  parameters describe a changing
environment of the local
system under consideration.
Often the relevant measured time series consists locally of
a Gaussian process, with the variance of those Gaussians
evolving on a longer time scale.
In nonequilibrium statistical mechanics, the technique
of superstatistics was introduced in
\cite{beck-cohen} and has since then provided a powerful tool
to describe a large variety of complex systems
for which there is change of environmental conditions
\cite{swinney,touchette,souza,chavanis,jizba,
frank,celia,straeten}.
A superstatistical
complex system is mathematically described as a
multi-scale superposition of two (or several)
statistics,
one corresponding to local equilibrium statistical mechanics (on a mesoscopic level modeled by a
linear Langevin equation leading to locally Gaussian behavior) and the other one corresponding to a slowly
varying parameter $a$ of the system.
Essential for this approach is the fact that there is sufficient time scale separation,
i.e. the local relaxation time of the system must be much shorter than the
typical time scale on which the parameter $a$ changes. There is interesting
mathematics associated with superstatistical complex systems. For example,
in a recent paper \cite{hanel}
Hanel, Thurner and Gell-Mann showed that the
superstatistical distribution function $f(a)$ cannot change in an
arbitrary way under macroscopic state changes of the system, and that the
underlying transformation group is the Euclidean group in 1 dimension.

In most applications in nonequilibrium statistical mechanics
the slowly varying control parameter $a$, underlying the superstatistical
dynamics,
is the local inverse temperature $\beta$ of the system, i.e. $a=\beta$.
However, in general
the control parameter $a$ can also have a different meaning.
For example in mathematical finance, $a$ is a vooatility parameter
describing changing market behavior and turbulences in the
market.
For a time series, $a$ is usually a local variance parameter that
can be directly extracted from suitably chosen time slices.
There are numerous interesting applications
of the superstatistics concept to real-world problems,
for example to
train delay statistics\cite{briggs},
hydrodynamic turbulence \cite{prl} and cancer survival statistics
\cite{chen}. For further applications, see
\cite{daniels,maya,reynolds,abul-magd,rapisarda,cosmic}.

In this paper,
we apply for the first time superstatistical techniques to
a complex systems of
solid state physics.
We are interested in the statistical
properties of currents flowing through
thin films when a voltage is applied.
Our measured system is a
thin aluminium-polymethylmethacrylate-aluminium film (Al-PMMA-Al).
A detailed experimental investigation of this system
was presented in \cite{hacinliyan2003, hacinliyan2006}.
A particular feature of this system is that when a voltage is applied,
the corresponding measured currents exhibit transient (nonstationary) behavior
with strong fluctuations. This can be formally associated
with transient chaotic behavior, and in the above papers formally a
small positive Liapunov exponent was extracted from the data.
More recently, Yalcin et al
\cite{singapore,yalcin} made a statistical analysis of the fluctuating
current data using $q$-statistics, inspired by the fact
that
$q$-statistics \cite{europhysics} is often
used for weakly chaotic systems which have approximately
zero Lyapunov exponent \cite{miritello, alessandro, tsallis-book}.

In this paper we show
that the measured current data can be well modelled by
a superstatistical dynamics.
However, in contrast to
previous superstatistical time series analysis
techniques described in \cite{straeten}, one has a typical
problem for these kinds of data: The time series is not long
enough to apply the standard
superstatistical techniques developed in \cite{straeten}.
Still, in this paper we show what one can typically do for
a short time series.
Our method described in the following is quite
simple and general and can be applied to other small data sets
in a similar way.

\section{Analyzing the experimental time series}

Figs.~1-3 show typical time series of increments
$I(t)=i(t+1)-i(t)$ of the measured current $i(t)$
through the Al-PMMA-Al film at three different temperatures.
The average value has been substracted, and all data are rescaled by dividing through
the standard deviation $\sigma$ determined from the entire time series.
We define $u(t)=(I(t)-\langle I(t) \rangle )/ \sigma$.

\

One clearly observes regions of strong activity interwoven with
periods of much calmer behavior, similar as for turbulent flows
of share price evolution data. To analyse these data, we
divide the time series into a couple of windows with
qualitatively similar behavior. These windows are shown
as vertical lines in Fig.1-3.


\begin{figure}
$\,$
\vspace*{2cm}

\epsfig{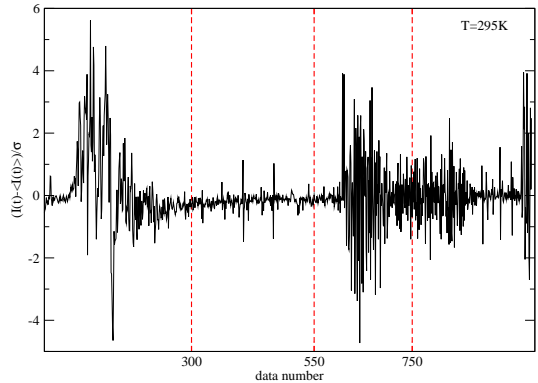}
\caption{The current magnitude differences
$I(t)=i(t+1)-i(t)$ for the transient current through thin Al-PMMA-Al film
at  $T=295K$  \cite{yalcin}}
\end{figure}

\begin{figure}
$\,$
\vspace*{2cm}

\epsfig {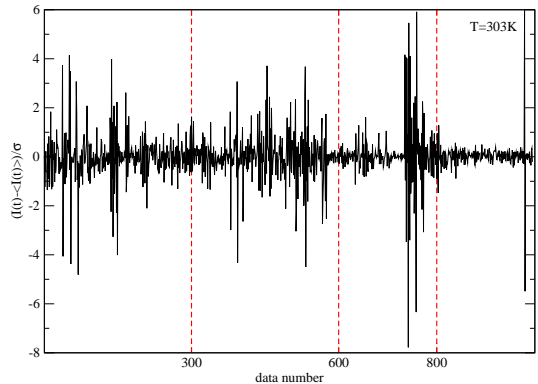}
\caption{The current magnitude differences
$I(t)=i(t+1)-i(t)$ for the transient current through thin Al-PMMA-Al film
at  $T=303K$}
\end{figure}

\begin{figure}
$\,$
\vspace*{2cm}

\epsfig {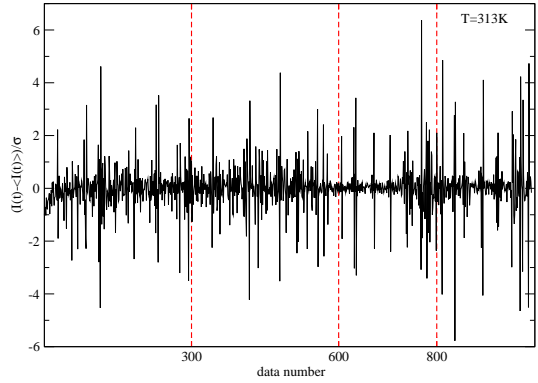}
\caption{The current magnitude differences
$I(t)=i(t+1)-i(t)$ for the transient current through thin Al-PMMA-Al film
at  $T=313K$}
\end{figure}

$\,$

The choice of the windows is rather arbitrary, they should
be large enough to contain enough statistics, but still be significantly
shorter than the entire time series.

For each local window, we look at a histogram of
the fluctuating currents. As shown in Fig.~4-7, locally
the behavior is well approximated by a Gaussian distribution
\begin{equation}
p(u|\beta)=\sqrt{\frac{\beta}{2\pi}}exp\left\{-\frac{1}{2}\beta u^{2}\right\}
\end{equation}
However, the local variance parameter $\beta$ varies strongly
from window to window and is thus itself a random variable,
as expected for superstatistical systems.


\begin{figure}
$\,$
\vspace*{2cm}

\epsfig{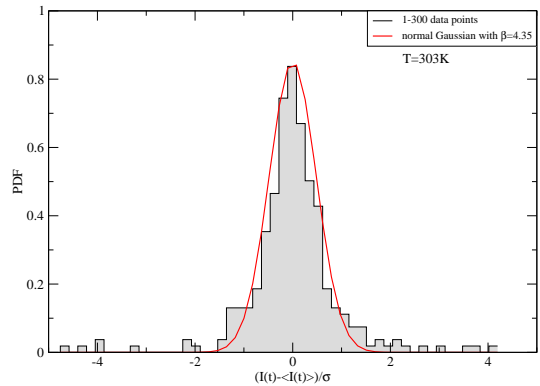}
\caption{Gaussian fit to the observed local probability density for data points 1-300 of the measured time series of
current differences $I(t)$ at $T=303K$.}
\end{figure}

\begin{figure}
$\,$
\vspace*{2cm}

\epsfig{file=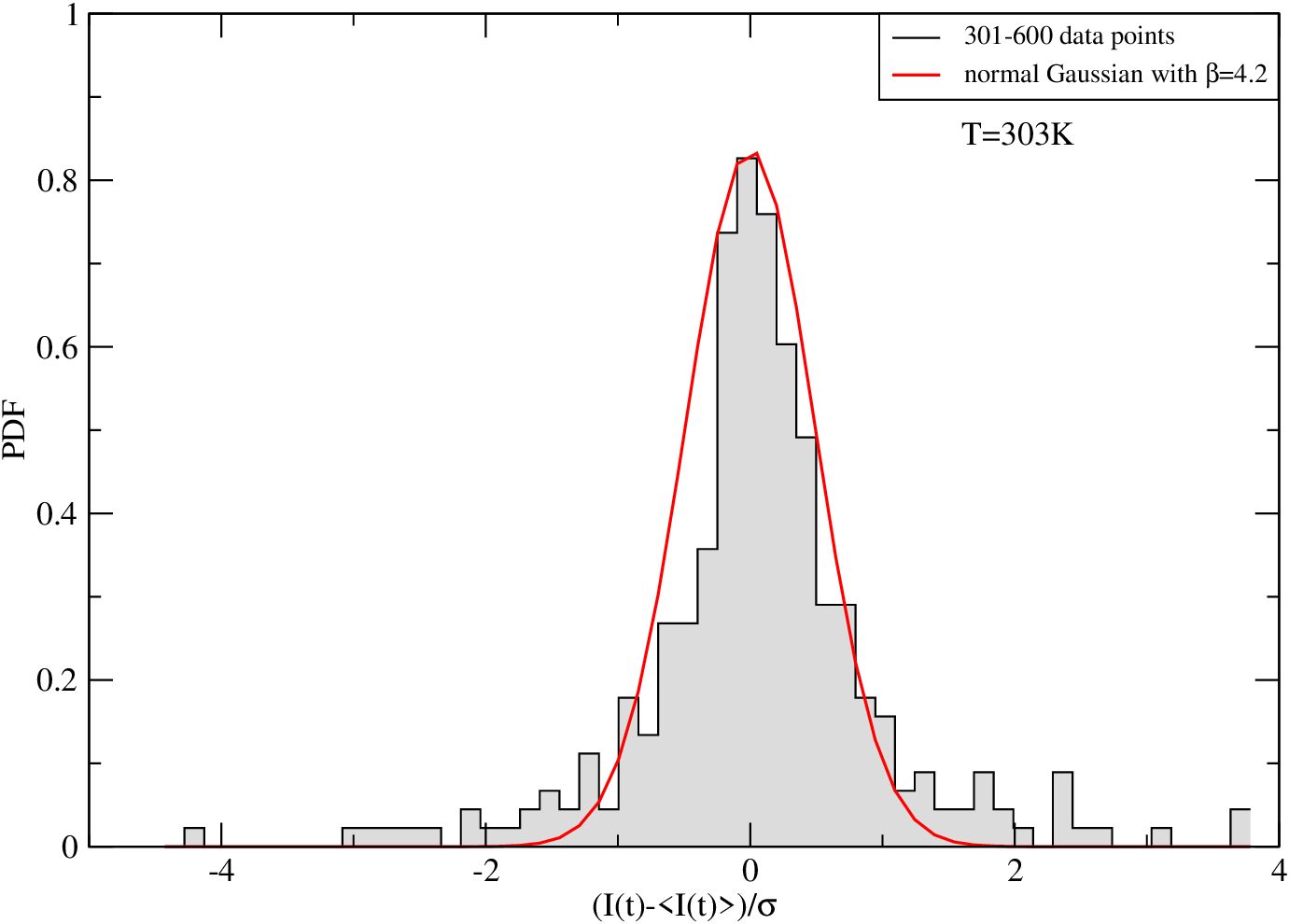, angle=0., width=7cm}
\caption{Same as Fig.~4 for data points 301-600 of the measured time series }
\end{figure}

$\,$

\begin{figure}
$\,$
\vspace*{2cm}

\epsfig{file=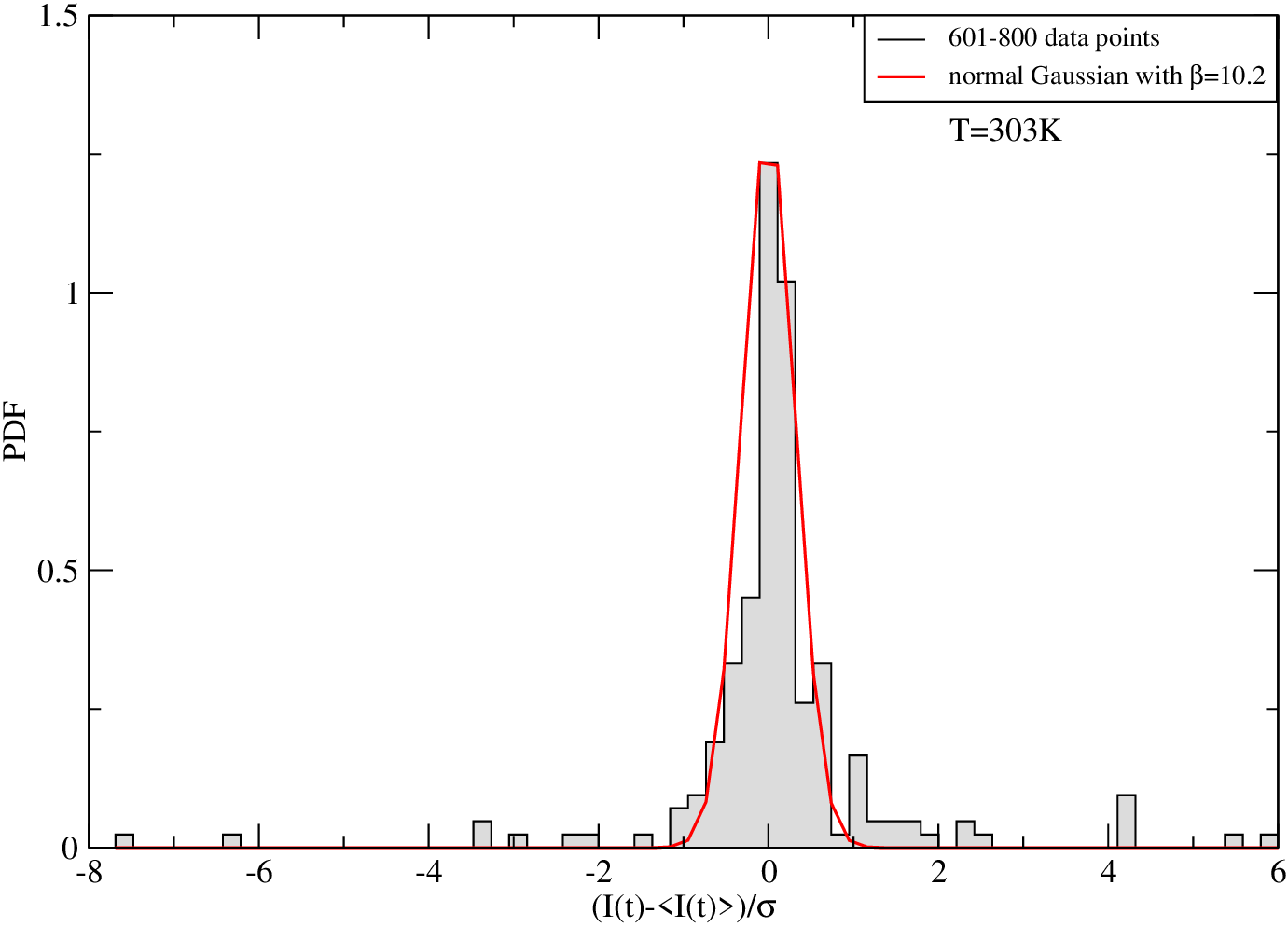, angle=0., width=7cm}
\caption{Same as Fig.~4 for 
 data points 601-800 of the measured time series}
 
\end{figure}

$\,$

\begin{figure}
$\,$
\vspace*{2cm}

\epsfig{file=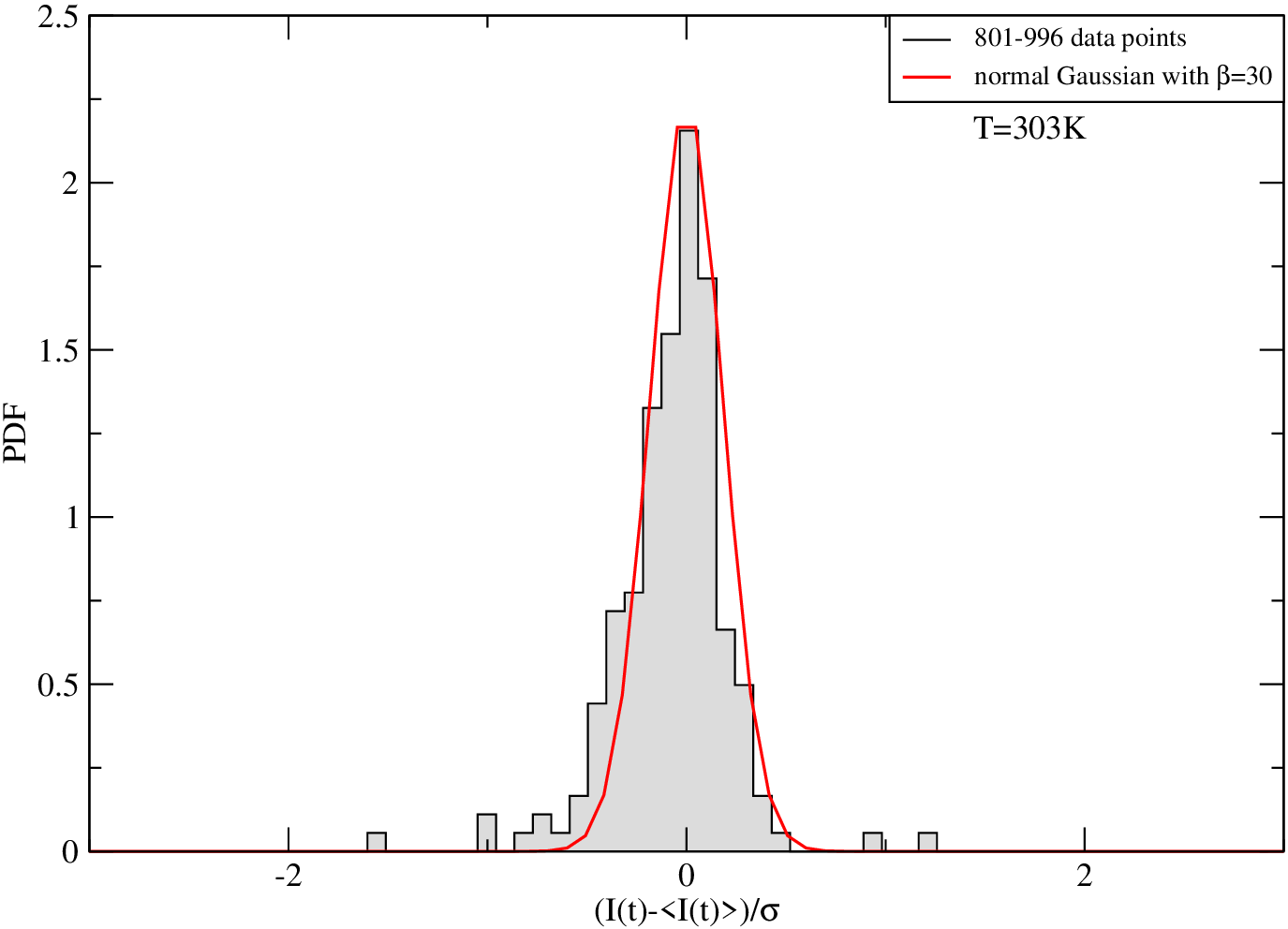, angle=0., width=7cm}
\caption{Same as Fig.~4 for 
 data points 801-996 of the measured time series}
 
\end{figure}

Tab.~1 shows the values of extracted variance parameters $\beta_i$
for the various windows and for different temperatures $T$ where the
experiment was performed.

We can only extract a few values $\beta_i$ from our short time series.
The general idea of superstatistics is that the marginal
distribution $p(u)$ is given as a superposition
\begin{equation}
p(u) =\int f(\beta) p(u|\beta) d\beta \label{aaa}
\end{equation}
for a suitable $f(\beta)$ describing the
distribution of $\beta$-values. For a short time series,
as in our case, one
can only make guesses of the relevant $f(\beta)$. For
this it is useful to look at the histogram of the entire
time series, sampling up the behavior in all windows. This is shown
in Fig.~8.


\begin{figure}
$\,$
\vspace*{2cm}

\epsfig{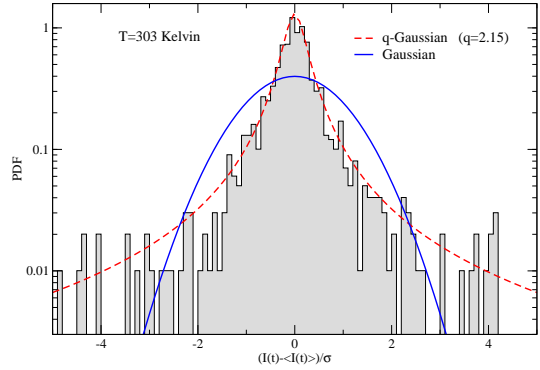}
\caption{Entire PDF of the current magnitude differences  $I(t)=i(t+1)-i(t)$ for the transient current through thin
Al-PMMA-Al at temperature $T=303K$ \ \cite{yalcin}}
\end{figure}

Apparently, the histogram data of the entire time series are well-fitted by $q$-Gaussians of the form
\begin{equation}
p(u) \sim (1+ \frac{1}{2}\tilde{\beta}(q-1)u^2)^\frac{1}{1-q}.   \label{bbb}
\end{equation}
The corresponding parameters $\tilde{\beta}$ and $q$ are shown
in Tab.1.

\

\begin {table}
\begin{center}
\begin{tabular}{|c|l|l|l|l|l|l|l|l|}
\hline
               & $\beta_{1} $ &$\beta_{2}$  &$\beta_{3}$  &$\beta_{4}$  &$\beta_{0}$  &$\tilde{\beta}$&$q$  \\\hline

T=295K &4.9    & 66      &8.8    &8.1   &21.95          &40  &2.3       \\\hline
T=303K &4.35  &4.2      &10.2  &30   &12.1875      &30  &2.15      \\\hline
T=313K &4        &6.95   &18.5  &6.2  &8.9125        &22  & 2.1         \\\hline
\end{tabular}
\caption{
The values of extracted variance parameters $\beta_{1} $,~$\beta_{2}$,~$\beta_{3}$,~$\beta_{4}$,and $\beta_{0}$ for the various slices, for the measured current through thin Al-PMMA-Al film at temperature $T=295,~303 ~and~ 313K$. Also shown $q$ and $\tilde{\beta}$.}
\end{center}
\end{table}

We cannot, as for longer time series \cite{straeten}, extract the relevant
distribution $f(\beta)$ from a histogram of $\beta_i$-values,
since there are only 4 data points. But
the fitting result (\ref{bbb}) can be used to theoretically
predict $f(\beta)$.  In our case
$f(\beta)$ is in good
approximation  a $\chi^2$ (or $\Gamma$)-distribution, meaning that if we were able
to look at much longer time series, then the $\beta_i$ would
be distributed according to

\begin{equation}
f(\beta)=\frac{1}{\Gamma(\frac{n}{2})}\left\{\frac{n}{2\beta_{0}}\right\}^{n/2}\beta^{n/2-1}
exp\left\{- \frac{n\beta}{2\beta_{0}}\right\} \label{ccc}
\end{equation}

This follows from doing the integral (\ref{aaa}) using the
distribution (\ref{ccc}), which directly leads to the
observed fitting result (\ref{bbb}).
The relation between the parameters $n$ and $q$ is \cite{prl2001}
\begin{equation}
q=1+\frac{2}{n+1}.
\end{equation}
Moreover, the parameter $\tilde{\beta}$
is related to the average value of $\beta$ by \cite{prl2001}
\begin{equation}
\beta_0 := \int f(\beta) \beta d\beta =  \frac{(3-q)}{2} \tilde{\beta} \label{ddd}
\end{equation}

The predicted superstatistical distributions $f(\beta)$ are plotted in Fig.~9
for the three different temperatures.
We can also
calculate the average value of $\beta_0$ directly from the time
series, namely
from the values $\beta_i$ fitted in the 4 windows, i.e.
$\beta_0=\frac{1}{4}(\beta_1+\beta_2+\beta_3+\beta_4)$. The result
is shown in Tab.~1 as well. This sum has huge error terms, because
only 4 windows enter, and the partioning into windows is
rather arbitrary.
Nevertheless, the obtained average values are
consistent with the theoretically predicted result (\ref{ddd})
within the statistical error bounds.


\begin{figure}
$\,$
\vspace*{2cm}

\epsfig{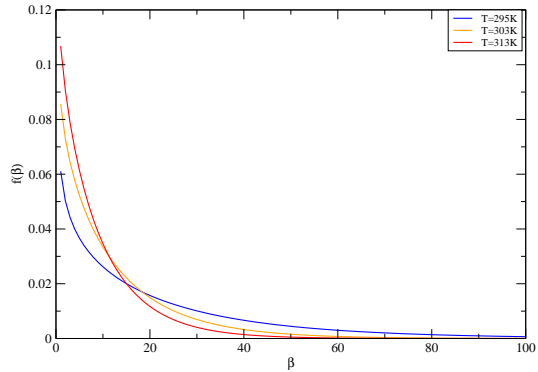}
\caption{$f(\beta)$ of the measured time series through Al-PMMA-Al film
at temperature $T=295,~303 ~and~ 313K$}
\end{figure}

To summarize,
for very long time series, one has theoretical methods to decide
whether a given time series is superstatistical or not \cite{straeten}.
For short time series, as in our case, these methods are not applicable.
Still one can do the method proposed in the current paper quite generally.
First one looks whether a partitioning into a few windows is possible
such that locally Gaussian behavior is observed. Then one determines
the local variance parameters $\beta_i$. From a histogram
of the entire time series one obtains a guess what the relevant
distribution $f(\beta)$ could be---in our case a $\chi^2$-distribution
but in general these can be other distributions such as
e.g. the lognormal or inverse-$\chi^2$ distribution \cite{swinney}.
After that guess, one can check consistency of
derived parameter relations within
the statistical error bounds, in our case given by relation (\ref{ddd})
but in general given by other relations, depending on $f(\beta)$.

\section{Acknowledgement}

G.C.Y. was supported by the Scientific Research Projects Coordination Unit of Istanbul University with project number 7441.
G.C.Y gratefully acknowledges the hospitality of Queen Mary University of London, School of Mathematical Sciences,
where this work was carried out. G.C.Y and C.B. would like to thank Prof. Murray Gell-Mann for useful discussions
on the subject.

\

\end{document}